\documentclass[rnaas]{aastex62}

\pdfoutput=1

\usepackage{lmodern}
\usepackage{microtype}
\usepackage{url}
\usepackage{amsmath}
\usepackage{amssymb}
\usepackage{natbib}
\usepackage{multirow}
\usepackage{graphicx}
\newcommand{\figureref}[1]{\ref{fig:#1}}
\newcommand{\Figure}[1]{Figure~\figureref{#1}}
\newcommand{\figurelabel}[1]{\label{fig:#1}}
\renewcommand{\eqref}[1]{\ref{eq:#1}}
\newcommand{\Eq}[1]{Equation~(\eqref{#1})}
\newcommand{\eq}[1]{\Eq{#1}}

\newcommand{\eqlabel}[1]{\label{eq:#1}}

\newcommand{\T}{\ensuremath{\mathrm{T}}}
\newcommand{\dd}{\ensuremath{ \mathrm{d}}}

\newcommand{\bvec}[1]{{\ensuremath{\boldsymbol{#1}}}}

\begin{document}\raggedbottom\sloppy\sloppypar\frenchspacing

\title{%
    Scalable backpropagation for Gaussian Processes using celerite
}

\author[0000-0002-9328-5652]{Daniel Foreman-Mackey}
\affil{Center for Computational Astrophysics, Flatiron Institute, New York, NY}


\section{Introduction}

This research note presents a derivation and implementation of efficient and
scalable gradient computations using the \emph{celerite} algorithm for Gaussian
Process (GP) modeling.
The algorithms are derived in a ``reverse accumulation'' or
``backpropagation'' framework and they can be easily integrated into existing
automatic differentiation frameworks to provide a scalable method for
evaluating the gradients of the GP likelihood with respect to all input
parameters.
The algorithm derived in this note uses less memory and is more efficient than
versions using automatic differentiation and the computational cost scales
linearly with the number of data points.

GPs \citep{Rasmussen:2006} are a class of models used
extensively in the astrophysical literature to model stochastic processes.
The applications are broad-ranging and some examples include the time domain
variability of astronomical sources \citep{Brewer:2009, Kelly:2014,
Haywood:2014, Rajpaul:2015, Foreman-Mackey:2017, Angus:2018}, data-driven
models of light curves or stellar spectra \citep{Wang:2012, Luger:2016,
Czekala:2017}, and the cosmic microwave background
\citep{Bond:1987,Wandelt:2003}.
In all of these applications, the calculation and optimization of the GP
marginalized likelihood function (here we follow the notation of
\citealt{Foreman-Mackey:2017})
\begin{eqnarray}\eqlabel{loglike}
\log \mathcal{L}(\bvec{\theta},\,\bvec{\alpha}) &=&
    -\frac{1}{2}\,\left[\bvec{y} - \bvec{\mu}_\bvec{\theta}\right]^\T\,
        {K_\bvec{\alpha}}^{-1}\,\left[\bvec{y}-\bvec{\mu}_\bvec{\theta}\right]
    -\frac{1}{2}\,\log\det K_\bvec{\alpha} + \mathrm{constant}
\end{eqnarray}
is generally the computational bottleneck.
The details of these models are omitted here (see \citealt{Rasmussen:2006} and
\citealt{Foreman-Mackey:2017} for details), but the key point is that, for a
dataset with $N$ data points, every evaluation of a GP model requires
computing the log-determinant and multiplying a vector by the inverse of
the $N \times N$ covariance matrix $K_\bvec{\alpha}$.
The computational cost of these operations scales as $\mathcal{O}(N^3)$ in the
general case, but the \emph{celerite} method was recently introduced in the
astronomical literature to compute the GP likelihood for a class of
one-dimensional models with $\mathcal{O}(N)$ scaling
\citep{Ambikasaran:2015, Foreman-Mackey:2017}.

The details of the \emph{celerite} method can be found in
\citet{Foreman-Mackey:2017} and I will not repeat them here.
The only difference in notation is that all the matrices in what follows are
the ``pre-conditioned'' matrices that are indicated with a tilde by
\citet{Foreman-Mackey:2017}.
The tilde is not included here for simplicity and to improve the readability
of the algorithms.
I will also use the symbol $P$ for the $(N-1) \times J$ pre-conditioning matrix
that was called $\phi$ by \citet{Foreman-Mackey:2017}.
The Cholesky factorization algorithm derived by \citet[][their
Equation~46]{Foreman-Mackey:2017} is as follows:

\medskip
\begin{minipage}{\linewidth}
\textbf{function} \texttt{celerite\_factor}($U$, $P$, $\bvec{d}$, $W$) \\
\hspace*{2em}\textsf{\# At input $\bvec{d} = \bvec{a}$ and $W = V$} \\
\hspace*{2em}$S \gets$ \texttt{zeros}($J$, $J$) \\
    \hspace*{2em}$\bvec{w}_1 \gets \bvec{w}_1 / d_{1}$ \\
\hspace*{2em}\textbf{for} $n = 2,\ldots,N$:\\
\hspace*{2em}\hspace*{2em}$S \gets \texttt{diag}(\bvec{p}_{n-1})\,[
    S + d_{n-1}\,{\bvec{w}_{n-1}}^\T\,{\bvec{w}_{n-1}}
]\,\texttt{diag}(\bvec{p}_{n-1})$ \\
\hspace*{2em}\hspace*{2em}$d_{n} \gets d_{n} - \bvec{u}_n\,S\,{\bvec{u}_n}^\T$\\
\hspace*{2em}\hspace*{2em}$\bvec{w}_n \gets \left[\bvec{w}_n -
    \bvec{u}_n\,S \right] / d_{n}$\\
    \hspace*{2em}\textbf{return} $\bvec{d}$, $W$, $S$
\end{minipage}
\medskip

\noindent In this algorithm, the \texttt{zeros}$(J,\,K)$ function creates a $J
\times K$ matrix of zeros, the \texttt{diag} function creates a diagonal
matrix from a vector, and $\bvec{x}_n$ indicates a row vector made from the
$n$-th row of the matrix $X$.
The computational cost of this algorithm scales as $\mathcal{O}(N\,J^2)$.
Using this factorization, the log-determinant of $K$ is
\begin{eqnarray}
\log \det K &=& \sum_{n=1}^N \log d_{n} \quad.
\end{eqnarray}
Similarly, \citet{Foreman-Mackey:2017} derived a $\mathcal{O}(N\,J)$ algorithm
to apply the inverse of $K$ (i.e.\ compute $Z = K^{-1}\,Y$) as follows
(Equations~47 and~48 in \citealt{Foreman-Mackey:2017}):

\medskip
\begin{minipage}{\linewidth}
\textbf{function} \texttt{celerite\_solve}($U$, $P$, $\bvec{d}$, $W$, $Z$) \\
\hspace*{2em}\textsf{\# At input $Z = Y$} \\
\hspace*{2em}$F \gets$ \texttt{zeros}($J$, $N_\mathrm{rhs}$) \\
\hspace*{2em}\textbf{for} $n = 2,\ldots,N$:\\
\hspace*{2em}\hspace*{2em}$F \gets \texttt{diag}(\bvec{p}_{n-1})\,[F +
    {\bvec{w}_{n-1}}^\T\,\bvec{z}_{n-1}]$\\
\hspace*{2em}\hspace*{2em}$\bvec{z}_n \gets \bvec{z}_n - \bvec{u}_n\,F$\\
\hspace*{2em}\textbf{for} $n = 1,\ldots,N$:\\
\hspace*{2em}\hspace*{2em}$\bvec{z}_n \gets \bvec{z}_n / d_{n}$\\
\hspace*{2em}$G \gets$ \texttt{zeros}($J$, $N_\mathrm{rhs}$) \\
\hspace*{2em}\textbf{for} $n = N-1,\ldots,1$:\\
\hspace*{2em}\hspace*{2em}$G \gets \texttt{diag}(\bvec{p}_{n})\,[G +
    {\bvec{u}_{n+1}}^\T\,\bvec{z}_{n+1}]$\\
    \hspace*{2em}\hspace*{2em}$\bvec{z}_n \gets \bvec{z}_n - \bvec{w}_n\,G$\\
\hspace*{2em}\textbf{return} $Z$, $F$, $G$
\end{minipage}
\medskip

\noindent The empirical scaling of these algorithms is shown in
\Figure{figure}.

\section{Gradients of GP models using celerite}

It is standard practice to make inferences using \Eq{loglike} by optimizing or
choosing a prior and sampling with respect to $\bvec{\theta}$ and
$\bvec{\alpha}$.
Many numerical inference methods (like non-linear optimization or Hamiltonian
Monte Carlo) can benefit from efficient calculation of the gradient of
\Eq{loglike} with respect to the parameters.
The standard method of computing these gradients uses the identity
\citep{Rasmussen:2006}
\begin{eqnarray}\eqlabel{naive-grad}
\frac{\dd \log \mathcal{L}}{\dd \alpha_k} &=&
    \frac{1}{2}\,\mathrm{Tr}\left[
        \left[
        \bvec{\tilde{r}}\,\bvec{\tilde{r}}^\T - {K_\bvec{\alpha}^{-1}}
        \right]
        \,\frac{\dd K}{\dd \alpha_k}
    \right]
\end{eqnarray}
where
\begin{eqnarray}
    \bvec{\tilde{r}} &=&
        {K_\bvec{\alpha}}^{-1}\,\left[\bvec{y}-\bvec{\mu}_\bvec{\theta}\right]
    \quad.
\end{eqnarray}
Similar equations exist for the parameters $\bvec{\theta}$.
Even with a scalable method of applying ${K_\bvec{\alpha}}^{-1}$, the
computational cost of \Eq{naive-grad} scales as $\mathcal{O}(N^2)$.
This scaling is prohibitive when applying the \emph{celerite} method to large
datasets and I have not found a simple analytic method of improving this
scaling for semi-separable matrices.
However, it was recently demonstrated that substantial computational gains can
be made by directly differentiating Cholesky factorization algorithms even in
the general case \citep{Murray:2016}.

Following this reasoning and using the notation from an excellent review of
matrix gradients \citep{Giles:2008}, I present the reverse-mode gradients of
the \emph{celerite} method.
While not yet popular within astrophysics, ``reverse accumulation'' of
gradients (also known as ``backpropagation'') has recently revolutionized the
field of machine learning \citep[see][for example]{LeCun:2015} by enabling the
non-linear optimization of models with large numbers of parameters.
The review \citep{Giles:2008} provides a thorough overview of these methods
and the interested reader is directed to that discussion for details and for
an explanation of the notation.

Using the notation from \citet{Giles:2008}, after some tedious
algebra, the reverse accumulation function corresponding to
\texttt{celerite\_factor} is found to be:

\medskip
\begin{minipage}{\linewidth}
\textbf{function} \texttt{celerite\_factor\_rev}($U$, $P$, $\bvec{d}$, $W$,
    $S$, $\bar{S}$, $\bar{\bvec{a}}$, $\bar{V}$) \\
\hspace*{2em}\textsf{\# At input $\bar{\bvec{a}} = \bar{\bvec{d}}$ and
    $\bar{V} = \bar{W}$}\\
\hspace*{2em}$\bar{U} \gets \mathrm{zeros}(N,\,J)$\\
\hspace*{2em}$\bar{P} \gets \mathrm{zeros}(N-1,\,J)$\\
\hspace*{2em}$\bar{\bvec{v}}_N \gets \bar{\bvec{v}}_N / d_N$\\
\hspace*{2em}\textbf{for} $n = N,\ldots,2$:\\
\hspace*{2em}\hspace*{2em}$\bar{a}_n \gets \bar{a}_n -
    \bvec{w}_n\,{\bar{\bvec{v}}_n}^\T$\\
\hspace*{2em}\hspace*{2em}$\bar{\bvec{u}}_n \gets - [\bar{\bvec{v}}_n +
    2\,\bar{a}_n\,\bvec{u}_n]\,S$\\
\hspace*{2em}\hspace*{2em}$\bar{S} \gets \bar{S} -
    {\bvec{u}_n}^\T\,[\bar{\bvec{v}}_n + \bar{a}_n\,\bvec{u}_n]$\\
\hspace*{2em}\hspace*{2em}$\bar{\bvec{p}}_{n-1} \gets \mathrm{diag}(\bar{S}\,S
    \,\mathrm{diag}(\bvec{p}_{n-1})^{-1} + \mathrm{diag}(\bvec{p}_{n-1})^{-1}\,
    S\,\bar{S})$\\
\hspace*{2em}\hspace*{2em}$\bar{S} \gets \mathrm{diag}(\bvec{p}_{n-1})\,
    \bar{S}\,\mathrm{diag}(\bvec{p}_{n-1})$\\
\hspace*{2em}\hspace*{2em}$\bar{d}_{n-1} \gets \bar{d}_{n-1} +
    \bvec{w}_{n-1}\,\bar{S}\,{\bvec{w}_{n-1}}^\T$\\
\hspace*{2em}\hspace*{2em}$\bar{\bvec{v}}_{n-1} \gets \bar{\bvec{v}}_{n-1}
    / d_{n-1} + \bvec{w}_{n-1}\,[\bar{S} + \bar{S}^\T]$\\
\hspace*{2em}\hspace*{2em}$S \gets \mathrm{diag}(\bvec{p}_{n-1})^{-1}\,S\,
    \mathrm{diag}(\bvec{p}_{n-1})^{-1}
    - d_{n-1}\,{\bvec{w}_{n-1}}^\T\,\bvec{w}_{n-1}$\\
\hspace*{2em}$\bar{a}_1 \gets \bar{a}_1 -
    \bar{\bvec{v}}_1\,{\bvec{w}_1}^\T$\\
\hspace*{2em}\textbf{return} $\bar{U}$, $\bar{P}$, $\bar{\bvec{a}}$, $\bar{V}$
\end{minipage}
\medskip

\noindent Similarly, the reverse accumulation function for to
\texttt{celerite\_solve} is:

\medskip
\begin{minipage}{\linewidth}
\textbf{function} \texttt{celerite\_solve\_rev}($U$, $P$, $\bvec{d}$, $W$,
    $Z$, $F$, $G$, $\bar{F}$, $\bar{G}$, $\bar{Y}$) \\
\hspace*{2em}\textsf{\# At input $\bar{Y} = \bar{Z}$}\\
\hspace*{2em}$\bar{U} \gets \mathrm{zeros}(N,\,J)$\\
\hspace*{2em}$\bar{P} \gets \mathrm{zeros}(N-1,\,J)$\\
\hspace*{2em}$\bar{\bvec{d}} \gets \mathrm{zeros}(N)$\\
\hspace*{2em}$\bar{W} \gets \mathrm{zeros}(N,\,J)$\\
\hspace*{2em}\textbf{for} $n = 1,\ldots,N-1$:\\
\hspace*{2em}\hspace*{2em}$\bar{\bvec{w}}_n \gets
    - \bar{\bvec{y}}_n\,G^\T$\\
\hspace*{2em}\hspace*{2em}$\bar{G} \gets \bar{G} - {\bvec{w}_n}^\T \,
    \bar{\bvec{y}}_n$\\
\hspace*{2em}\hspace*{2em}$\bvec{z}_n \gets \bvec{z}_n + \bvec{w}_n\,G$\\
\hspace*{2em}\hspace*{2em}$G \gets \mathrm{diag}(\bvec{p}_n)^{-1}\,G$\\
\hspace*{2em}\hspace*{2em}$\bar{\bvec{p}}_n \gets \mathrm{diag}(
    \bar{G}\,G^\T)$\\
\hspace*{2em}\hspace*{2em}$\bar{G} \gets
    \mathrm{diag}(\bvec{p}_n)\,\bar{G}$\\
\hspace*{2em}\hspace*{2em}$G \gets G - {\bvec{u}_{n+1}}^\T\,\bvec{z}_{n+1}$\\
\hspace*{2em}\hspace*{2em}$\bar{\bvec{u}}_{n+1} \gets
    \bvec{z}_{n+1}\,\bar{G}^\T$\\
\hspace*{2em}\hspace*{2em}$\bar{\bvec{y}}_{n+1} \gets
    \bvec{u}_{n+1}\,\bar{G}$\\
\hspace*{2em}\textbf{for} $n = 1,\ldots,N$:\\
\hspace*{2em}\hspace*{2em}$\bar{\bvec{y}}_n \gets \bar{\bvec{y}}_n/d_n$\\
\hspace*{2em}\hspace*{2em}$\bar{d}_n \gets -
    \bvec{z}_n\,{\bar{\bvec{y}}_n}^\T$\\
\hspace*{2em}\textbf{for} $n = N,\ldots,2$:\\
\hspace*{2em}\hspace*{2em}$\bar{\bvec{u}}_n \gets \bar{\bvec{u}}_n
    - \bar{\bvec{y}}_n\,F^\T$\\
\hspace*{2em}\hspace*{2em}$\bar{F} \gets \bar{F} - {\bvec{u}_n}^\T \,
    \bar{\bvec{y}}_n$\\
\hspace*{2em}\hspace*{2em}$F \gets \mathrm{diag}(\bvec{p}_{n-1})^{-1}\,F$\\
\hspace*{2em}\hspace*{2em}$\bar{\bvec{p}}_{n-1} \gets \bar{\bvec{p}}_{n-1} +
    \mathrm{diag}(\bar{F}\,F^\T)$\\
\hspace*{2em}\hspace*{2em}$\bar{F} \gets
    \mathrm{diag}(\bvec{p}_{n-1})\,\bar{F}$\\
\hspace*{2em}\hspace*{2em}$F \gets F - {\bvec{w}_{n-1}}^\T\,\bvec{z}_{n-1}$\\
\hspace*{2em}\hspace*{2em}$\bar{\bvec{w}}_{n-1} \gets \bar{\bvec{w}}_{n-1} +
    \bvec{z}_{n-1}\,\bar{F}^\T$\\
\hspace*{2em}\hspace*{2em}$\bar{\bvec{y}}_{n-1} \gets \bar{\bvec{y}}_{n-1} +
    \bvec{w}_{n-1}\,\bar{F}$\\
\hspace*{2em}\textbf{return} $\bar{U}$, $\bar{P}$, $\bar{\bvec{d}}$,
    $\bar{W}$, $\bar{Y}$
\end{minipage}
\medskip

\noindent A reference C++ implementation of this algorithm can be found online
\citep{Foreman-Mackey:2018} and \Figure{figure} shows the performance of this
implementation.

\newpage
\section{Discussion}

This research note presents the algorithms needed to efficiently compute
gradients of GP models applied to large datasets using the \emph{celerite}
method.
These developments increase the performance of inference methods based on
\emph{celerite} and improve the convergence properties of non-linear
optimization routines.
Furthermore, the derivation of reverse accumulation algorithms for
\emph{celerite} allow its integration into popular model building and
automatic differentiation libraries like Stan \citep{Carpenter:2015},
TensorFlow \citep{Abadi:2016}, and others.

\acknowledgments
It is a pleasure to thank
Eric Agol,
Sivaram Ambikasaran, and
Victor Minden
for conversations that inspired this work.
A reference implementation of these algorithms, benchmarks, and tests can be
found at \url{https://github.com/dfm/celerite-grad} and
\citet{Foreman-Mackey:2018}.

\begin{figure}[htbp]
\begin{center}
\includegraphics[width=0.8\textwidth]{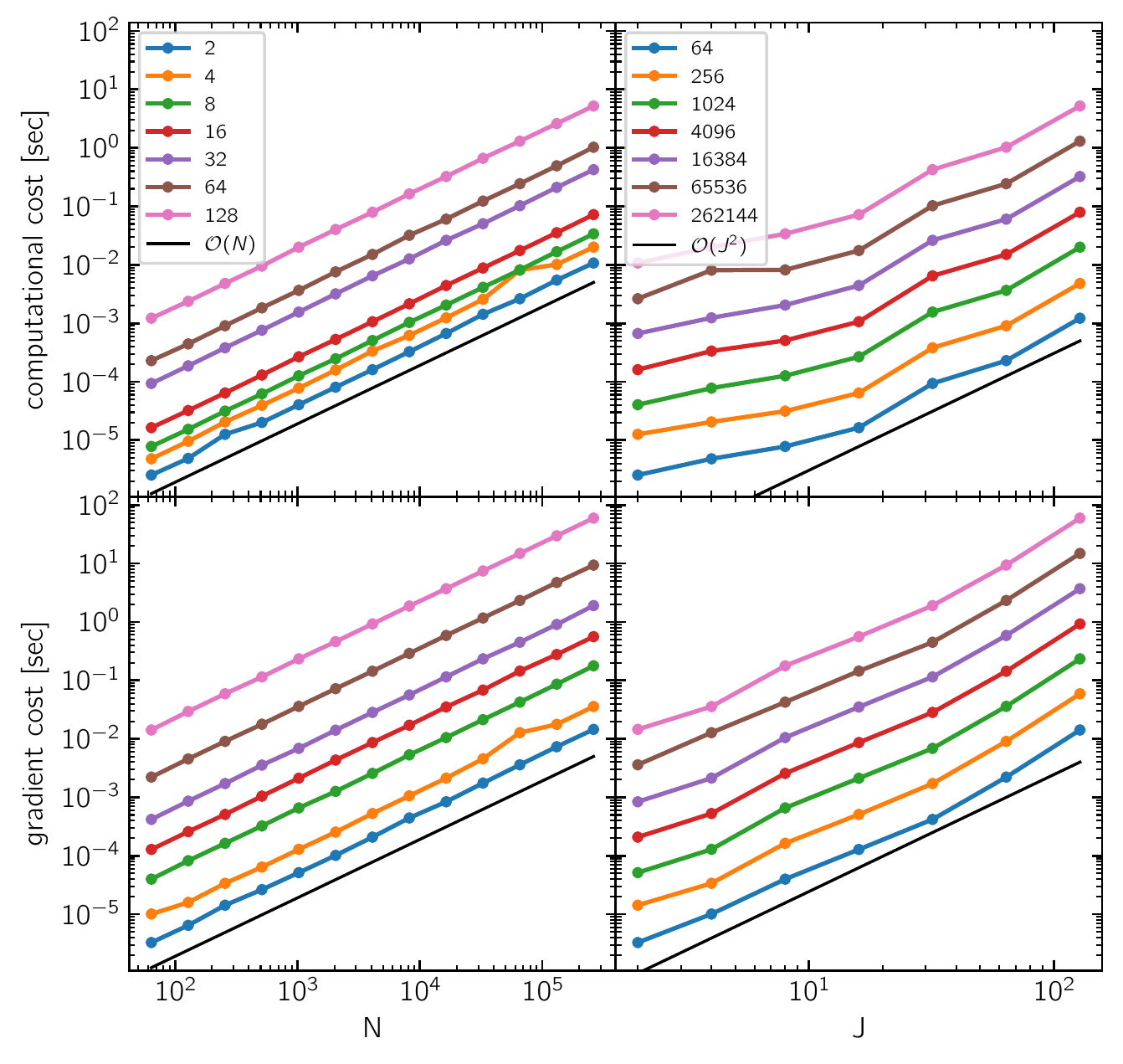}
\caption{%
    The empirical computational scaling of the algorithms presented in this
    note.
    \emph{(top row)}: The cost of computing \eq{loglike} as a
    function of the number of data points ($N$; left) and the model complexity
    ($J$; right).
    \emph{(bottom row)}: The cost of computing the gradient of \eq{loglike}
    with respect to the vector $\bvec{a}$ and the matrices $U$, $V$, and $P$.
    In the left panels, each line corresponds to a different value of $J$ as
    indicated in the legend (with $J$ increasing from bottom to top).
    Similarly, in the right panels, the lines correspond to different values
    of $N$ increasing from bottom to top.
    In both cases, the theoretical scaling is $\mathcal{O}(N,J^2)$.
\figurelabel{figure}}
\end{center}
\end{figure}

\bibliography{celerite}

\end{document}